\documentclass[final,3p,times,twocolumn]{elsarticle}
\usepackage{amssymb}
\usepackage{url}
\biboptions{sort&compress}

\journal{Physics Letters B}

\begin{document}

\begin{frontmatter}



\title{Systematic reduction of the proton-removal cross section in neutron-rich medium-mass nuclei}


\author[USC]{J. D{\'i}az-Cort\'es}
%\cortext[cor1]{Corresponding author.}
%\ead{david.loureiro@usc.es}
\author[USC]{J. Benlliure}
\author[USC]{J.L. Rodr{\'i}guez-S\'anchez}
\author[USC]{H. \'Alvarez-Pol}
\author[TU]{T. Aumann}
\author[TAM]{C.A. Bertulani}
\author[CENBG]{B. Blank}
\author[USC]{E. Casarejos\fnref{label1}}
\fntext[label1]{Present address: University of Vigo, E-36200  Vigo, Spain}
\author[USC]{D. Cortina-Gil}
\author[USC]{D. Dragosavac}
\author[GSI]{V. F\"ohr}
\author[INFN]{A. Gargano}
\author[USC]{M. Gasc\'on}
%\fntext[label2]{Present address: Stanford University, CA. 94305, USA}
\author[CSW]{W. Gawlikowicz}
\author[CHA]{A. Heinz}
\author[UH]{K. Helariutta}
\author[GSI]{A. Keli\'c-Heil}
\author[GSI]{S. Luki\'c}
%\fntext[label4]{Present address: Karlsruhe Institute of Technology, D-76021 Karlsruhe, Germany }
\author[GSI]{F. Montes\fnref{label5}}
\fntext[label5]{Present address: NSCL/MSU, East Lansing, MI. 48824, USA.}
\author[USC]{D. P\'erez-Loureiro\fnref{label6}}
\fntext[label6]{Present address: Canadian Nuclear Laboratories Ltd, Chalk River Laboratories, Chalk River, Canada K0J 1P0} 
\author[UW]{L. Pie\'nkowski}
\author[GSI]{K-H. Schmidt}
\author[GSI]{M. Staniou}
\author[VINCA]{K. Suboti\'c}
\author[GSI]{K. S\"ummerer}
\author[CEA]{J. Taieb}
\author[UW]{A. Trzci\'nska}

\address[USC]{IGFAE, Universidade de Santiago de Compostela, E-15782 Spain}
\address[TU] {Institut f\"ur Kernphysik, Technische Universit\"at Darmstadt, 64289 Darmstadt, Germany}
\address[TAM] {Texas A\&M University-Commerce, 75428 Commerce, Texas, United States of America}
\address[CENBG]{Centre d'Etudes Nucleaires, F-33175 Bordeaux-Gradignan Cedex, France}
\address[GSI]{GSI Helmholtzzentrum  f\"ur Schwerionenforschung, D-64291 Darmstadt, Germany}
\address[INFN] {Istituto Nazionale di Fisica Nucleare, Complesso Universitario di Monte S. Angelo, Via Cintia, I-80126 Napoli, Italy}
\address[UW]{Heavy Ion Laboratory, University of Warsaw, PL-02-093 Warsaw, Poland}
\address[CSW]{Cardinal Stefan Wyszynski University, PL-01-938  Warsaw, Poland}
\address[CHA]{Chalmers University of Technology, SE-41296 Gothenburg, Sweden}
\address[UH]{University of Helsinki, FI-00014 Helsinki, Finland}
\address[VINCA]{Institute of Nuclear Sciences Vin\v{c}a, University of Belgrade, 11001 Belgrade, Serbia}
\address[CEA]{CEA, DAM, DIF F-91297 Arpajon, France}

\begin{abstract}
Single neutron- and proton-removal cross sections have been systematically measured for 72 medium-mass neutron-rich nuclei around Z=50 and energies around 900$A$ MeV using the FRagment Separator (FRS) at GSI. Neutron-removal cross sections are described by considering the knock-out process together with initial- and final-state interactions. Proton-removal cross sections are, however, significantly smaller than predicted by the same calculations. The observed difference can be explained as due to the knockout of short-correlated protons in neutron-proton dominating pairs.
\end{abstract}

\begin{keyword}
Nucleon removal \sep Medium-mass neutron-rich nuclei \sep Short-range correlations

\end{keyword}
\end{frontmatter}

% \linenumbers

\section{Introduction}
Single-nucleon knockout at intermediate and high energies is a widely used tool to investigate the structure of the atomic nucleus as the removed nucleon is expected to provide information on the previously occupied single-particle state. This success relies both, on the well defined experimental conditions to investigate these reactions, but also because Eikonal models provide a relatively simple description of the reaction. Following these ideas, direct kinematics ($e,e'p$) low-momentum transfer reactions on stable nuclei measured in the 80s and 90s probed the limits of the independent particle shell-model picture. Those investigations showed a 30-40\% reduction in the spectroscopic factors quantifying the shell-model fragmentation of the occupancy of single-particle states around the Fermi level \cite{Lapikas:1993,Kelly:1996}. Less than half of this reduction has been attributed to long-range correlations responsible for collective nuclear modes \cite{Dickhoff:2004}. Similar investigations using proton beams have provided complementary information on the single-particle structure of stable nuclei \cite{Wakasa:2017}.

High momentum-transfer proton knockout experiments induced by electrons \cite{Niyazov:2004} and protons \cite{Tang:2003} probed the existence of short-lived correlated nucleon pairs by the identification of knockout nucleons forming pairs with relatively low center-of-mass momentum ($k_{tot}<k_F$) but rather large relative momentum ($k_{rel}>k_F$). Those short-range correlated (SRC) nucleon pairs, produced by the short-range components (scalar or tensor) of the nuclear force, involve around 20\% of the nucleons in a given nucleus \cite{Egiyan:2006},  explaining the depopulation of single-particle states below the Fermi momentum ($k_F$) and the population of higher energy states. Recently, it has also being shown that 80\% of the SRC pairs are neutron-proton pairs \cite{Subedi:2008}, indicating the predominance of the tensor interaction. As consequence, the number of protons in SRC pairs increases with neutron-excess in the nucleus \cite{Hen:2014,Duer:2018}. 

Nucleon-induced knockout reactions in inverse kinematics opened the possibility to investigate the structure of unstable fast-moving secondary projectile nuclei \cite{Kobayashi:1988}. Moreover, exclusive single-nucleon knockout measurements including $\gamma$-ray detection have provided spectroscopic information on the ground state of the investigated nuclei \cite{Navin:1998}. This technique has also been used more recently to investigate the reduction of the single-particle spectroscopic strength around the Fermi level by introducing a ''quenching'' factor defined as the ratio between measured single-nucleon knockout cross sections and calculated ones, taking into account shell-model spectroscopic factors \cite{Gade:2004}. 

The systematic investigation of single-nucleon knockout reactions on light projectile nuclei covering a large range in neutron excess and energies up to around 100$A$ MeV, not only showed a sizable quenching of the expected spectroscopic strength but also a strong dependence with the neutron excess. Indeed, it was shown that the removal of loosely bound nucleons induces a small or negligible quenching of the spectroscopic strength while the quenching for the removal of deeply bound nucleons is beyond 50\% \cite{Gade:2008}. However, transfer \cite{Flavigny:2013} and quasi-free, ($p,2p$) or ($p,pn$), nucleon removal \cite{Atar:2018,Gomez:2018} do not show any clear dependence of the quenching of the spectroscopic strength with the neutron excess for the same nuclei.

The puzzling dependence of the spectroscopic strength quenching with the neutron excess, could be qualitatively understood taking into consideration the predominance of neutron-proton SRC pairs in nuclei. The knockout of a SRC proton from a neutron-rich nucleus would cause the counterpart nucleon, mostly a neutron, to recoil and be ejected as well \cite{Subedi:2008}. Therefore, the increase of the number of SRC protons with the neutron excess \cite{Duer:2018}, will reduce, accordingly, the probability of single-proton removal processes. To contribute to this discussion, in this work we have performed a systematic investigation of single-proton and single-neutron removal reactions for medium-mass nuclei, around Z=50, at high energies.

\section{Experiment and measurements}
The experiment was performed at the GSI facility in Darmstadt taking advantage of two different beams impinging a 1 g/cm$^2$ beryllium target to produce large isotopic chains of medium-mass nuclei. Fission of 950$A$ MeV $^{238}$U projectiles, produced neutron-rich nuclei around $^{132}$Sn and fragmentation of 1200$A$ MeV $^{132}$Xe projectiles leading to less neutron-rich nuclei around Z=50. As the reactions were induced in inverse kinematics and high energies, the forward focused reaction fragments were analyze with the FRS magnetic spectrometer \cite{Geissel:1992}.

In this experiment the two sections of the FRS were tuned as a two independent magnetic spectrometers. The first section was used to separate the medium-mass nuclei produced in the target located at the entrance of the FRS. The identification of the transmitted nuclei was achieved by measuring the magnetic rigidity (B$\rho$), time of flight (ToF), and energy loss ($\Delta$E) by using fast plastic scintillators located behind the first and second dipoles of the FRS, and time-projection chambers (TPCs) and a fast ionization chamber placed at the intermediate-image plane. An additional 2591 mg/cm$^2$ beryllium target was placed at the intermediate-image plane to induce nucleon removal reactions. Those reaction residues were separated using the second section of the FRS and identified by the same B$\rho$-ToF-$\Delta$E method. Additional details on the experiment can be found in Ref. \cite{Perez:2011}. 

% measuring the ToF from the intermediate- to the final-image planes with plastic scintillators and TPC and ionization chambers located at the final-image plane for tracking and $\Delta$E measurements. 

\begin{center}
\begin{figure}[h]
\begin{center}
\includegraphics[width=0.45\textwidth]{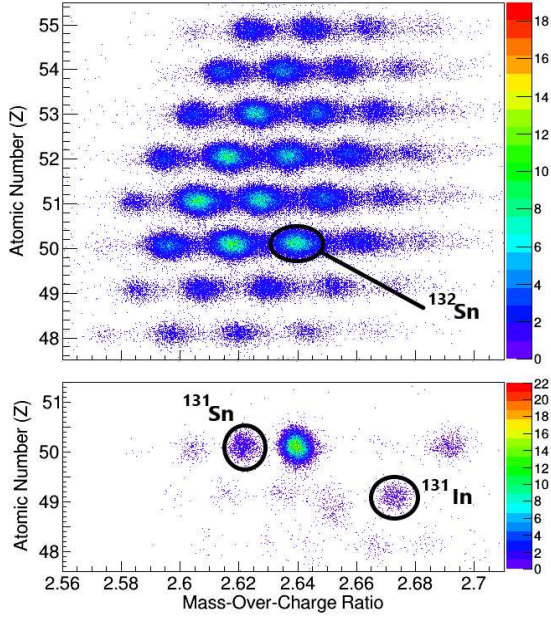}
\caption{Identification matrices of the incoming cocktail beam obtained with the first section of the FRS for a magnetic setting centered on $^{132}$Sn (upper panel) and of the single-proton ($^{131}In$) and single-neutron ($^{131}$Sn) residual fragments identified with the second section of the FRS (lower panel).}
\label{fig1}
\end{center}
\end{figure}
\end{center}

Figure \ref{fig1} shows the identification matrices obtained with the $^{238}$U beam for a magnetic tuning of the first section of the FRS centered on $^{132}$Sn (upper panel) and the combined tunings of the second section centered on $^{131}$Sn and $^{132}$In (lower panel). The resolution (FWHM) achieved in the isotopic identification of the fragments transmitted through the first section of the FRS was $\Delta$Z/Z$\approx$2.6 10$^{-3}$, $\Delta$A/A$\approx$1.2 10$^{-3}$, and in the second section $\Delta$Z/Z$\approx$3.0 10$^{-3}$, $\Delta$A/A$\approx$7.8 10$^{-4}$.

To investigate long isotopic chains of medium-mass nuclei around Z=50 several magnetic tunings of the first section of the FRS were used. Those tunings were centered on $^{119}$Sn, $^{124}$Sn and $^{126}$Sn with the $^{132}$Xe beam, and  $^{128}$Sn, $^{130}$Sn, $^{132}$Sn and $^{136}$Sn with the $^{238}$U beam. For each of this tunings the second section of the FRS was centered on the corresponding one-neutron and one-proton removal residues. With these measurements we were capable to determine the one-neutron removal cross section for 72 isotopes of tellurium, antimony, tin and indium and the proton-removal cross sections for 13 isotopes of the same elements.

\begin{center}
\begin{figure}[h]
\begin{center}
\includegraphics[width=0.5\textwidth]{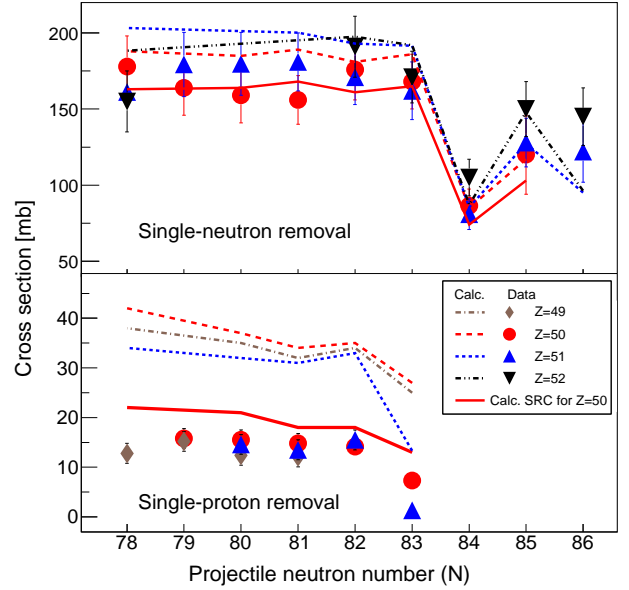}
\caption{One-neutron (upper panel) and one-proton removal cross sections measured in this work for different isotopes of indium, tin, antimony, and tellurium. Lines represent model prediction described in the text.}
\label{fig2}
\end{center}
\end{figure}
\end{center}

Figure \ref{fig2} shows the single-neutron removal cross sections (upper panel), and the single-proton removal cross sections (lower panel), measured in this work as function of the neutron number of the initial nucleus. In both cases, the measured cross sections are rather similar for the isotones of the three elements covered by the measurements. The single-neutron removal cross sections are also similar for all isotopes from $N$=78 until $N$=83. At $N$=84 all cross sections drop significantly, increasing again at $N$=85. The observed decrease of the cross sections at $N$=84 may be explained by the fact that the excited states populated in the residual nuclei lie above the the neutron separation. For $N$=84 projectiles, the single-neutron removal produces with a large probability $N$=83 residues with a hole states in the 1$d_{3/2}$ or 0$h_{11/2}$ orbitals, below the shell gap, and one neutron occupying the 1$f_{7/2}$ orbital above the $N$=82 shell gap. Those configurations correspond to excitations energies in the residual nuclei larger than the $N$=82 gap energy, $\sim$3.7 MeV. Because of these large excitation energy values, and the low neutron binding energies ($S_n$) of the $N$=83 residues (e.g. $S_n$($^{133}$Sn)=2.4 MeV), the survival probability against neutron emission of the single-neutron removal residues from $N$=84 projectiles will be rather small. Conversely, $N$=82 remnants produced in the single neutron-removal of $N$=83 projectiles do not populate orbitals above the $N$=82 shell gap and the hole states will be close to the Fermi level. The excitation energy gained in that case will be rather small, while binding relatively large (e.g. $S_n$($^{132}$Sn)=7.2 MeV). The survival probability of the remnants produced in the single-neutron removal of $N$=83 isotopes will be then rather high. This argument is also valid for all $N<84$ projectile nuclei. The increase of the cross section for $N=85$ can also be explained by the large binding energies as compare to the excitation energies of the populated states. 

Single-proton removal cross sections, shown in the lower panel in Fig. \ref{fig2}, are also very similar for all isotones until $N$=83 where they drop. Moreover, the cross section for $^{133}$Sn is slightly higher than for $^{134}$Sb. This behavior can also be explained by using similar arguments. The decrease in cross section at $N$=83 is due to the lower survival probability of those remnants against neutron emission because of the lower binding energies. Moreover, the larger excitation energy expected by the removal of a proton in nuclei with occupied orbitals above the $Z$=50 shell reduces the survival probability of $^{134}$Sb remnants respect to $^{133}$Sn ones. Below $N=83$ the effect of the larger excitation energies in $Z>50$ remnants does not seem to be sufficient to overcome the larger neutron binding energies. 

Another remarkable fact is, that single-proton removal cross sections are in about an order of magnitude smaller than single-neutron removal ones. This difference has also been observed in nucleon-removal cross sections obtained with stable medium-mass and heavy nuclei (e.g. $^{112,124}$Sn \cite{Rodriguez:2017}, $^{136}$Xe \cite{Benlliure:2008}, $^{197}$Au \cite{Rejmund:2001}, $^{208}$Pb \cite{Audouin:2006} and $^{238}$U \cite{Taieb:2003})), and with few unstable medium-mass nuclei (e.g. $^{132}$Sn \cite{Perez:2011}, $^{90}$Sr \cite{Wang:2016}, $^{137}$Cs \cite{Wang:2016}). The common issue to all these measurements is that they concern neutron-rich nuclei. One could then expect that the mentioned difference in cross sections could be due to the larger excess of neutrons at the nuclear periphery. However, previous works, taking into account realistic radial distributions of protons and neutrons in some of these nuclei, could not account for the observed differences in cross section \cite{Rodriguez:2017}.

\begin{center}
\begin{figure}[ht]
\begin{center}
\includegraphics[width=0.45\textwidth]{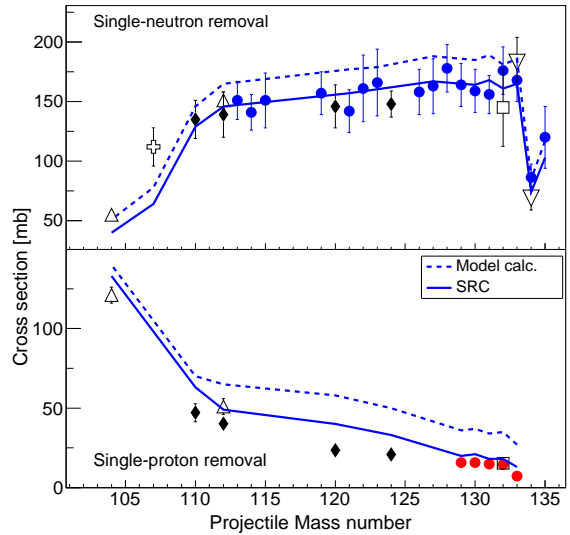}
\caption{Single-neutron (upper panel) and single-proton (lower panel) removal cross sections for different tin isotopes measured in this work (dots) and by other authors. Lines represent model calculations described in the text.}
\label{fig3}
\end{center}
\end{figure}
\end{center}

In fig. \ref{fig3} we depict the single-neutron (upper panel) and single-proton (lower panel) removal cross sections measured in this work for different tin isotopes (dots) together with similar measurements reported in literature ($^{133,134}$Sn \cite{Vaquero:2017} (inverted triangles), $^{132}$Sn \cite{Perez:2011} (square), $^{124,120,112,110}$Sn \cite{Rodriguez:2017} (romboids),$^{112,104}$Sn \cite{Audirac:2013} (triangles), and $^{107}$Sn \cite{Cerizza:2016} (cross)). The good agreement between coincident measurements validate the results obtained in the present work. Moreover, with the complete set of existing data we systematically cover the single-neutron and single-proton removal cross sections for a large fraction of tin isotopes between $^{104}$Sn and $^{134}$Sn.

\section{Model calculations}

The modelization of inclusive nucleon-removal reactions requires the description of all processes leading to a final A-1 residual nucleus. This includes the direct knockout of a single nucleon producing a final residue in a bound state, but also initial- an final-state interactions that may contribute to nuclear excitations feeding the A-1 channel, or to produce A-1 unbound remnants. In particular, inelastic electromagnetic and nuclear excitations, or re-scattering of the knockout nucleons should be considered. Moreover, one needs an accurate description of the structural properties of the involved nuclei, as their radial distributions of protons and neutrons, and the corresponding single-particle states and spectroscopic factors describing its ground state. 

The absence of a complete theory requires to couple consistently, models describing the reaction and the structural properties of the involved nuclei.  Because of the additional complexity to describe initial- and final state interactions, we decided to use an advanced intra-nuclear cascade (INC) model providing an accurate description of the knockout processes, including realistic radial profiles for protons and neutrons \cite{Rodriguez:2017b}. Moreover, this formalism provides a good description of final-state interactions of the knockout nucleons, as well as nucleon excitations present in the energy range used in this work. 

Particle-hole excitations in the remnants were computed using experimental information, when available, and realistic shell model calculations providing the energies and occupation of single-particle orbitals. These calculations where done using a $^{88}$Sr core with the orbitals 1p$_{1/2}$, 0g$_{9/2}$, 0g$_{7/2}$, and 1d$_{5/2}$ as model space for protons, and 0g$_{7/2}$, 1d$_{5/2}$, 1d$_{3/2}$, 2s$_{1/2}$, and 0h$_{11/2}$ for neutrons \cite{Coraggio:2016,Gargano:2017}. Hole states were randomly defined considering the occupation of the different valence orbitals and the overlap between the corresponding wave functions and the range of impact parameters for knockout processes provided by the INC model.

The final excitation of the knockout remnants was obtained by adding to particle-hole excitations the energy gained in the re-scattering of the outgoing nucleons as computed by the intra-nuclear cascade model. The survival probability of these remnants was computed as the fraction of the excitation function below the particle emission threshold. 

Cross sections and energy gained by electromagnetic and nuclear excitations due to the isovector giant-dipole and the isoscalar giant-quadrupole resonances of the projectile nuclei were computed according to Ref. \cite{Bertulani:2003}. In this case, the fraction of the excitation function between one and two nucleon emission thresholds was accounted as part of the single nucleon-removal cross section. Because most of the investigated nuclei are neutron-rich, these excitations mostly contributes to the single-neutron removal cross section.

\section{Results and discussion}

\begin{center}
\begin{figure}[ht]
\begin{center}
\includegraphics[width=0.5\textwidth]{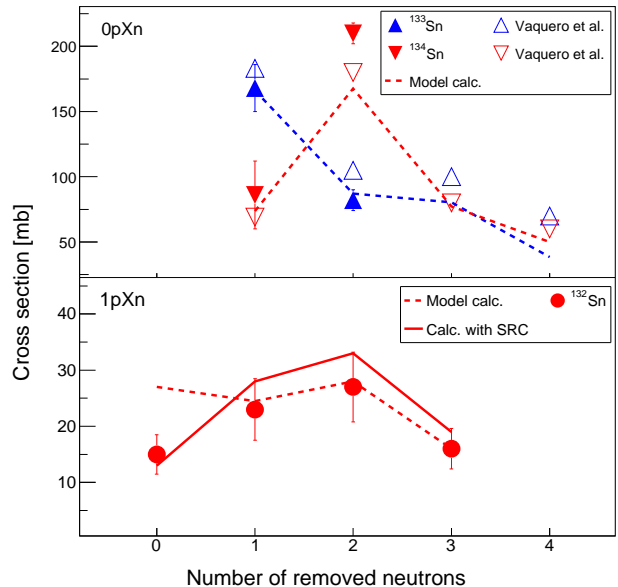}
\caption{Upper panel: 0$p$x$n$ removal cross sections for $^{133}$Sn and $^{134}$Sn compared to previous measurements by Vaquero et al. \cite{Vaquero:2017}. Lower panel: 1$p$x$n$ removal cross sections for $^{132}$Sn. Lines represent model predictions described in the text.}
\label{fig4}
\end{center}
\end{figure}
\end{center}

The results of the model calculations are depicted with dashed lines in Figs. \ref{fig2}, \ref{fig3} and \ref{fig4}. The calculations provide a rather good description of the measured neutron-removal cross sections (see upper panel in Fig. \ref{fig2}), in particular the little difference between isotopic chains and the drop in cross section at N=84, due to the N=82 shell gap, as explained in Sect 2. For heavier isotones, the model calculations slightly underestimate the measured cross sections. This deviation can be explained by the recently observed gamma decay of unbound states in these nuclei \cite{Vaquero:2017}, not considered in our calculations. The single knockout of lighter isotopes measured in other works, shown in the upper panel of Fig. \ref{fig3}, is also nicely described. 

To further check the calculations, in the upper panel of Fig. \ref{fig4} we depict the multi-neutron removal cross sections for $^{133}$Sn and $^{134}$Sn. The opposite behavior observed in the cross section for the one- and two-neutron removal for the two nuclei is explained by the influence of the shell gap in the one-neutron removal for N=84 isotones. Again, the model calculations provide an accurate description of the multi-nucleon removal processes. This result is important because it confirms that the model not only properly describes the excitation energy gained by the knockout remnants around the nucleon emission thresholds, mostly determined by particle-hole excitations, but also the higher excitations energies induced by initial- and final-state interactions. 

The surprising result is the large over-prediction of the proton-removal cross sections obtained with the same model calculations, as shown by the dashed lines in the lower panels of Figs. \ref{fig2} and \ref{fig3}. Moreover, the one-proton $x$-neutron (1$pxn$) removal cross sections, depicted in the lower panel of Fig. \ref{fig4}, show that this over-prediction mostly affects the single-proton removal channel. 

The fact that some 20\% of the nucleons inside a nucleus belong to SRC neutron-proton pairs may provide a rather simple explanation \cite{Duer:2018}. The knockout of SRC protons induces the emission of the neutron partner, because of their large relative momentum, depopulating the 1$p$0$n$ channel in favor of the 1$p$1$n$. Moreover, in neutron-rich systems the relative fraction of protons in SRC pairs is rather large (i.e. in $^{132}$Sn 13 protons and 13 neutrons belong to SRC pairs, representing 26\% or the protons and 16\% of the neutrons).

The same model calculations presented above, but including the presence of 20\% SRC neutron-proton pairs depopulating the 1$p$0$n$ channel in favor of the 1$p$1$n$, provide a rather satisfactory description of the measured one-proton removal cross sections, as shown by the solid lines in the lower panels of Figs. \ref{fig2}, \ref{fig3} and \ref{fig4}. The unobserved increase in the cross section for the 1$p$1$n$ (lower panel in Fig.4) is explained by the excitation energy gained in the rescattering of the two emitted nucleons. On the other hand, the relative smaller presence of neutrons in SRC pairs only produces a small reduction in the previously calculated one-neutron removal cross sections within our experimental uncertainties, as shown by the solid lines in the upper panels of the same figures.

\section{Conclusions}
Measurements of the single-neutron and single-proton removal cross sections over long isotopic chains of medium-mass nuclei confirm the systematic reduction of the proton-removal cross sections when compared to model calculations describing the neutron-removal process. Similar reductions in these cross sections were previously observed in lighter and heavier nuclei, indicating that this is a general feature. The presence of SRC on nucleon pairs, and more particularly the dominance of neutron-proton pairs, provides a satisfactory explanation. The removal of a SRC proton mostly populates the 1$p$x$n$ channels rather than the 1$p$0$n$. This effect is even larger in neutron-rich systems with a larger relative presence of protons in SRC pairs.

\section*{Acknowlegments}
This work was partially funded by the Spanish Ministry for Science and Innovation under grant FPA2007-6252, the programme ``Ingenio 2010, Consolider CPAN'', 2010/57 ``Grupos de referencia competitiva Xunta de Galicia'' and EC under the  EURISOL-DS contract No. 515768 RIDS. 


%% References


\bibliographystyle{model1-num-names}
\bibliography{bibliography}
%%\nocite{*}


\end{document}